\documentclass[twoside,fleqn,espcrc2]{article}

\makeatletter

\providecommand{\LyX}{L\kern-.1667em\lower.25em\hbox{Y}\kern-.125emX\@}

\newcommand{\AmS}{{\protect\the\textfont2
  A\kern-.1667em\lower.5ex\hbox{M}\kern-.125emS}}
\makeatother

\begin{document}

\title{Gravity coupling from micro-black holes}

\author{Fabio Scardigli}
\date{}
\maketitle
Institute for Theoretical Physics, University of Bern, 

Sidlerstrasse 5, 3012 Bern, Switzerland.

e-mail: fabio@itp.unibe.ch

\begin{abstract}
Recently much work has been done in lowering the Planck threshold of quantum
gravitational effects (sub-millimeter dimension(s), Horava-Witten fifth dimension,
strings or branes low energy effects, etc.). Working in the framework of 4-dim
gravity, with semi-classical considerations based on Hawking evaporation of
planckian micro-black holes, I shall show here as quantum gravity effects could
occur also near GUT energies.
\end{abstract}

\section{Introduction}

As it is well known, the standard picture of supersymmetric GUT models predicts
that the coupling of the three gauge interactions (electromagnetic, weak and
color) unify to a good accuracy at an energy around \( 2\cdot 10^{16} \) GeV.
Gravity, on the contrary, presents another natural energy scale: quantum fluctuations
of the gravitational field seem to become important only when we observe them
at the Planck length \( L_{P}=(G\hbar /c^{3})^{1/2}=1.6\cdot 10^{-33} \) cm.
Energy fluctuations of the order of \( E_{P}=\frac{1}{2}(\hbar c^{5}/G)^{1/2}=6\cdot 10^{18} \)
GeV create at this scale microblack holes that modify the topology of the spacetime.
At the same scale, following the common views, gravity coupling (that is the
\char`\"{}strenght\char`\"{} of gravitational interaction) should become comparable
with those of the other three gauge interactions.\\
 During the last years, studies based on string theories have changed this vision
~\cite{a}. In the early times of superstring theory, it was usual to associate
it with (sub)planckian physics, because the theory provided an ultraviolet regulator
of quantum gravity. More recently, a number of authors have considered the possibility
that the compactification energy scale is far lower, with a fundamental scale
of string theory being as low as TeV. Since the gauge couplings are inversely
proportional to the volume of compactification space, this implies large compactification
volumes and therefore large extradimension(s). By modifying the compactification
radius one can tune the couplings of gauge interactions, including gravity.\\
 In the \char`\"{}brane-world\char`\"{} picture the gauge interactions are localized
on p-branes (with p \( \leq  \) 9) while gravity propagates in different spacetime
dimensions. Precisely, gauge fields and particles with gauge charges move only
on the \char`\"{}walls\char`\"{} (i.e. on p-branes) while gravity moves also
in the bulk, the spacetime region whose the p-branes are the boundary ~\cite{b}.
In the Horava-Witten proposal the branes represent 3+1 dimensional walls in
which all the standard model particles live, while gravity moves in the 4+1
dimensional bulk between the walls. The other six additional dimensions of string
theory should be much smaller than that inhabited by gravity. The wall is then
9+1 dimensional in all and the complete spacetime is 10+1 dimensional. One can
choose the size of the 11th dimension (the one inhabited only by gravity) so
that the gravity coupling meets those of the other three forces at the same
common scale, presumably the GUT scale ~\cite{c}. And the couplings of the
other interactions remain untouched by this extra dimension.

\section{Gravity coupling from micro-black holes}

The purpose of this section is to show how the threshold of quantum gravitational
effects could be considerable lowered from \( M_{Planck} \) to almost \( M_{GUT} \),
\emph{without} the use of large extra dimensions, of 11 dimensional M-Theory
or of Horava-Witten model. Working in the framework of semiclassical 4-dim gravity,
a careful analysis brings to the conclusion that quantum gravity effects could
be present also at the GUT scale.\\
 Let us begin by reminding the process of creation of a planckian micro-black
hole. We can use the Heisenberg inequality \( \Delta p\Delta x\geq \hbar /2 \)
casted in the form \( \Delta E\Delta x\geq \hbar c/2 \) (because \( \Delta E\simeq c\Delta p \)
in our high energy situation) to observe that in a space region of width \( \Delta x \)
the metric field can fluctuate with an amplitude in energy of \( \Delta E\simeq \hbar c/2\Delta x \).
If the gravitational radius \( R_{g}=2G\Delta E/c^{4} \) associated with the
energy \( \Delta E \) equals the width \( \Delta x \) of the space region,
a micro-black hole originates ~\cite{d}. This happens when \( \Delta x=L_{P} \)
and the typical energy of the process is of course the Planck energy \( E_{P} \).
Planckian microholes are therefore typical quantum gravitational objects.\\
 We want now to calculate the lifetime of one of these microholes. This can
be done in at least three different ways. The first is to follow the pure thermodynamical
approach of the semiclassical Hawking decay ~\cite{Hawking}. Of course many
criticisms can be raised towards this approach, mainly because the extremely
high energy situation seems to forbid any semiclassical calculus and even the
use of the notion of temperature. The second method is to follow a microcanonical
approach ~\cite{Casadio}, which, using the microcanonical ensamble, avoids
the problematic use of the concept of \char`\"{}temperature\char`\"{} at these
very high energies. The third way is to make use of the semiclassical Hawking
decay, but corrected with opportune factors in order to account for the great
number of decay channels (i.e. species of particles) that the microhole can
radiate into during the final stages of its (short!) life.\\
 Although the microcanonical approach seem to be the more appropriate for this
high energy situation, it is based on an exponentially rising density of states,
which is interpreted by considering black holes as extended quantum objects
(\emph{p}-branes). We want on the contrary to work in a completely non-stringy
environment and to remain in the realm of standard model and 4-dim gravity,
in order to test how non-stringy physics can tackle the problem of gravity coupling.
Therefore we choose here the third method of calculus.\\
 Let us start with the thermodynamical approach and introduce after the \char`\"{}multi-channels\char`\"{}
correction factor. The lifetime of a black hole of given initial mass \( M_{0} \),
loosing mass via Hawking radiation can be calculated in a simple manner. Using
the Stefan-Boltzmann law

\begin{equation}
W=\sigma T^{4}
\end{equation}

we can say that the variation \( dM \) of the mass of the hole in a time \( dt \)
is

\begin{equation}
dM=-\frac{\sigma T^{4}S}{c^{2}}dt
\end{equation}

where \( S \) is the surface of the hole and \( T \) is the Hawking temperature.
For a Schwarzschild black hole we have

\begin{equation}
S=4\pi R_{g}^{2}
\end{equation}

\[
with\; \; \; R_{g}=\frac{2GM}{c^{2}}\; \; \; and\; \; \; T=\frac{\hbar c^{3}}{8\pi kGM}.\]

The equation in \( dM \) becomes

\begin{equation}
dM=-\frac{\lambda }{M^{2}}dt
\end{equation}

with 
\begin{equation}
\lambda =\frac{\sigma \hbar ^{4}c^{6}}{2^{8}\pi ^{3}G^{2}k^{4}}\simeq 3.93\cdot 10^{24}\; g^{3}s^{-1}.
\end{equation}

This gives the mass of the hole at time \( t \)

\begin{equation}
M(t)=(M_{0}^{3}-3\lambda t)^{1/3}
\end{equation}

and therefore the lifetime of the hole is

\begin{equation}
t_{0}=\frac{M_{0}^{3}}{3\lambda }=\frac{2^{8}\pi ^{3}G^{2}k^{4}}{\sigma \hbar ^{4}c^{6}}\frac{M_{0}^{3}}{3}.
\end{equation}

In particular for a micro-black hole of a Planck mass, \( M_{0}=M_{P}=E_{P}/c^{2} \),
we obtain for the lifetime

\begin{equation}
t_{0P}=\frac{2^{5}}{3}\frac{\pi ^{3}k^{4}}{\sigma }\frac{1}{\hbar ^{3}c^{2}}(\frac{G\hbar }{c^{5}})^{1/2}\simeq 2.01\cdot 10^{3}\tau _{P}
\end{equation}

where \( \tau _{P}=(G\hbar /c^{5})^{1/2} \) is the Planck time. (Remember that
\( E_{P}\tau _{P}=\hbar /2 \)).\\
 The precendent derivation, making use of the Stefan-Boltzmann law, takes into
account only the pure electromagnetic black body emission (i.e. photons). In
order to consider also other particles species, we have now to introduce a \char`\"{}multi-channels\char`\"{}
correction factor, \( f_{c} \). The equation (4) now reads

\begin{equation}
dM=-\frac{\lambda f_{c}}{M^{2}}dt
\end{equation}

where \( f_{c} \) accounts for the degrees of freedom of each emitted particle
specie contributing to the energy loss. Various forms have been proposed for
\( f_{c} \) ~\cite{MacGibbon}. In the standard model picture \( f_{c} \)
never exceeds a value of order \( 10^{2} \) (and also in a supersymmetric picture
\( f_{c} \) is again of this order of magnitude). The equation (8) hence becomes

\begin{equation}
t_{0P}\simeq \frac{2.01\cdot 10^{3}}{f_{c}}\; \tau _{P}.
\end{equation}

We note now that, in absence of gravitational effects, an energy fluctuation
of the same size of a planckian micro hole would have a lifetime of just only
one Planck time (if we take into account the Heisenberg principle only). The
decay time is slowed down by the presence of the event horizon of the micro
hole, which traps the energy and emits it only at a slow rate via Hawking evaporation.
Thus we can consider the quantum micro hole, which has a lifetime of the order
of \( 20\tau _{P} \), as a sort of \emph{metastable quantum state}. From the
theory of the decay of metastable states ~\cite{Landau} we can infer that the
decay probability \( dP \) of a state during the time interval \( (t,t+dt) \)
is proportional to

\begin{equation}
exp[-\frac{\Gamma }{\hbar }t]
\end{equation}

where \( \Gamma  \) is the width in energy of the state. The mean lifetime
of the state is

\begin{equation}
\tau =\frac{\hbar }{\Gamma }.
\end{equation}

We know that the lifetime of the metastable micro hole quantum state is \( t_{0P} \)
and from here we can calculate its width in energy

\begin{equation}
\Gamma _{0P}=\frac{\hbar }{t_{0P}}=\frac{2\cdot E_{P}f_{c}}{2.01\cdot 10^{3}}=6\cdot 10^{15}f_{c}\; GeV.
\end{equation}

For \( f_{c} \) of the order \( 10^{2} \) we obtain \( \Gamma _{0P}=6\cdot 10^{17} \)
GeV. This width is so huge that should allow the existence of planckian micro-black
holes also at the GUT threshold or very near to it, quite below the Planck threshold.\\
 Incidentally, it is also interesting to note that, by reversing the argument,
we can obtain a valuation of the number of particles species present in Nature.
In fact, if we demand that

\begin{equation}
\Gamma _{0P}=E_{P}-E_{GUT}
\end{equation}

we get then \( f_{c}\simeq 996 \).

\section{Conclusion}

The GUT energy scale seems to emerge in a natural way from a careful analysis
of the micro-black hole metastable quantum state. This seems to indicate that
typical quantum gravity objects can be present also at energies well below the
Planck threshold, and this result is obtained in a completely non-stringy framework.
The quantum gravity scale could be fixed at the GUT scale (almost) and no more
at the Planck scale. This agrees also with the simplest inflationary potential
invoked to explain the density fluctuations as measured by COBE (see Banks \( et\; al. \)
~\cite{a}).

\end{document}